\documentclass[
    ,final            
  ]
  {aipproc}

\layoutstyle{6x9}

\begin{document}

\title{Finite-Size Scaling from the non-perturbative Renormalization Group}


\author{Bertram Klein}{
  address={Physik Department, Technische Universit\"at M\"unchen, James-Franck-Strasse, 85747 Garching, Germany}
}

\author{Jens Braun}{
  address={Theory Group, TRIUMF, 4004 Wesbrook Mall, Vancouver, BC V6T 2A3, Canada
}
}

\begin{abstract}
The phase diagram of QCD at finite temperature and density and the existence of a critical point are currently very actively researched topics.
Although tremendous progress has been made, in the case of two light quark flavors even the order of the phase transition at zero density is still under discussion.
Finite-size scaling is a powerful method for the analysis of phase transitions in lattice QCD simulations. From the scaling behavior, critical exponents can be tested and the order as well as the universality class of a phase transition can be established. This requires knowledge of the critical exponents and the scaling behavior.
We use a non-perturbative Renormalization Group method to obtain critical exponents and the finite-size scaling functions for the O(4) universality class in three dimensions. These results are useful for a comparison to the actual scaling behavior in lattice QCD simulations with two flavors, as well as for an estimate of the size of the scaling region and the deviations from the expected scaling behavior. 
\end{abstract}

\maketitle


\section{Introduction}
In the application of lattice gauge theory to the QCD phase diagram, considerable effort has been spent on its successful extension to finite quark density. Nevertheless, even at vanishing quark density, there remains some doubt about the exact nature of the phase transition at finite temperature for two light quark flavors. While there is a lot of evidence that the phase transition is of second order \cite{Aoki:2006we}, results by Di Giacomo and collaborators with staggered fermions indicate that it might be a first-order transition \cite{D'Elia:2005bv,Cossu:2007mn}.

It is difficult to establish the order of a phase transition from a simulation in a finite volume: The partition function remains an analytical function of the thermodynamic parameters, no singular behavior appears 
and strictly speaking no phase transition occurs. Dynamical breaking of a continuous symmetry does not occur, either, and the analysis is further complicated by the necessary explicit breaking of such a symmetry. Therefore a careful finite-size scaling analysis is an important tool to determine the order and universality class of the transition. 

In QCD with two massless quark flavors, the SU(2) $\times$ SU(2) chiral flavor symmetry of the Lagrangian is spontaneously broken to SU(2) in the vacuum. 
Assuming a second-order transition to restore this symmetry, for massless quarks one expects this transition to fall into the O(4) universality class. In the case of staggered fermions on the lattice, the symmetry is further reduced and one expects an O(2) transition. This is expected if the QCD phase transition is dominated by the restoration of chiral symmetry.

Results with a modified chiral lattice action ($\chi$QCD) indicate that current simulation volumes might still be too small and therefore outside of the finite-size scaling region \cite{Kogut:2006gt}. In particular, this is relevant for the case of an O(2) symmetry, where lattice spin model calculations show that the scaling region is narrow. This could account for cases where no second-order scaling in either the O(2) or the O(4) universality class is observed.

In this context, it is very useful to analyze the scaling behavior in lattice QCD by comparing it to the behavior of O(N) models.
So far, mainly lattice spin models have been used for this purpose \cite{Engels:2001bq,Schulze:2001zg}.
In the current contribution, we present Renormalization Group results for the O(4) universality class in $d=3$ dimensions. An advantage of this method is that we can obtain results over a very wide parameter range, which in turn allows for very direct comparisons to lattice results.
 
\section{Renormalization Group method}
We employ a non-perturbative Renormalization Group (RG) method to calculate thermodynamic quantities. It includes long-range fluctuations and is thus capable of describing critical behavior. A review of functional RG methods can be found in e.g.  \cite{Berges:2000ew,Pawlowski:2005xe,Gies:2006wv}.

The RG approach is formulated in terms of a scale-dependent effective action which includes quantum fluctuations between an infrared cutoff scale $k$ and a UV cutoff $\Lambda$.
The change of the potential under a change of the cutoff scale $k$ is governed by a flow equation. 
For a specific choice of RG scheme \cite{Litim:2001up,Litim:2001hk, Bohr:2000gp}, the RG flow equation for the effective potential for the O(N) model in a $d$-dimensional infinite volume is given by
 \begin{eqnarray}
 k \frac{\partial}{\partial k} U_k( \sigma, \vec{\pi}) &=&   \frac{(k^2)^{d/2+1}}{(4 \pi)^{d/2}}\frac{1}{\Gamma(d/2+1)} \left( \frac{(N-1)}{k^2 + M^2_\pi(k)} + \frac{1}{k^2 +M^2_\sigma(k)}   \right).
 \end{eqnarray}
In finite volume, we use particular properties of our RG scheme to simplify sums over the discrete momenta \cite{Braun:2005fj}. 
A study of the RG scheme dependence of our results in finite volume is in preparation.
We expand the 
effective 
potential in local 
$n$-point 
interactions around the vacuum expectation value $\sigma_0(k)$:
\[
U_k(\sigma, \pi) = a_0(k) + a_1(k) (\sigma^2 + \vec{\pi}^2 -\sigma_0^2(k)) + a_2(k) (\sigma^2 + \vec{\pi}^2 -\sigma_0^2(k))^2 + \ldots - H \sigma.
\]
The condition $2 a_1(k) \sigma_0(k)= H $ ensures that the minimum is at $(\sigma, \vec{\pi}) = (\sigma_0(k), \vec{0})$. 

The RG flow equation is solved numerically; input to the calculation are the values of the couplings at the UV scale $\Lambda$. For $d=3$, the initial value for the minimum $\sigma_0(\Lambda)$ serves as a proxy for the temperature, $(\sigma_0(\Lambda) - \sigma_0^{\mathrm{critical}}(\Lambda)) \sim (T-T_c)$.
In the present case, we have chosen a cutoff scale of $\Lambda = 1.0$ GeV, which is of the order of the lattice cutoff ($\pi/a \approx 1.5$ GeV) in a typical thermodynamic lattice calculation (with $a\approx 0.2 - 0.3$ fm) and thus appears to be a reasonable choice for a first comparison. Ultimately, this should be adjusted to match specific lattice results for a comparison.

\section{Scaling in Infinite Volume}

Critical points, such as a second-order phase transition, are characterized by a diverging correlation length $\xi$. The associated critical long-range fluctuations lead to universal behavior where certain quantities are independent of the details of the system. Accordingly, systems can be grouped into \emph{universality classes}.
Close to the critical point, the behavior is then characterized by a small number of critical exponents specific to the universality class. 
\begin{table}
\begin{tabular}{llllll}
\hline
  &
  & \tablehead{1}{c}{b}{$\nu$}
  & \tablehead{1}{c}{b}{$\beta$}
  & \tablehead{1}{c}{b}{$\eta$}
  & \tablehead{1}{c}{b}{$\delta$}   \\
\hline
J. Engels {\it et al.} \cite{Engels:2001bq} & lattice & 0.7423 & 0.380  & 0.024
\tablenote{value calculated with scaling relations from the other exponents} 
& 4.86\\
D. Litim, and J. M. Pawlowski \cite{Litim:2001hk} & RG& 0.8043 & 0.4022$^*$ & $-$ & 5.00$^*$ \\
our work & RG & 0.8053(6) & 0.4030(3) & 0.0046(4)$^*$  & 4.9727(5)\\
\hline
\end{tabular}
\caption{\label{tab:critex}Results for critical exponents for O(4) in $d=3$ from RG and lattice calculations}
\end{table}
We obtain critical exponents $\beta, \nu, \delta$ directly from fits to the observables $M=\sigma_0$ and $1/\xi = M_\sigma$. The results are given in Tab.~\ref{tab:critex} and are in complete agreement with the RG fixed point analysis in \cite{Litim:2001hk}, and in reasonably good agreement with spin model lattice results.  Deviations are most likely due to the restriction to local couplings.

\begin{figure}
  \includegraphics[height=.2\textheight]{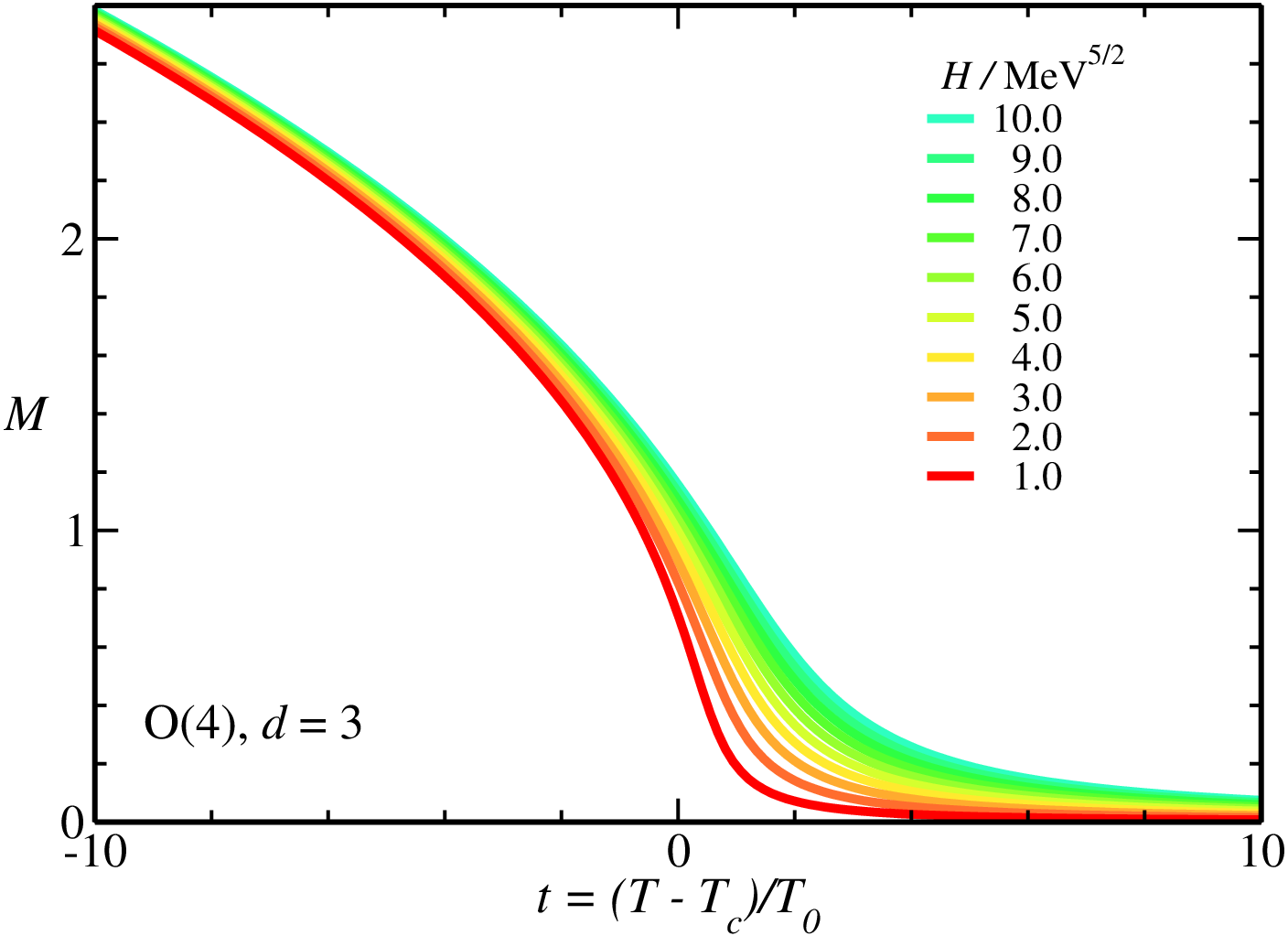}
   \includegraphics[height=.2\textheight]{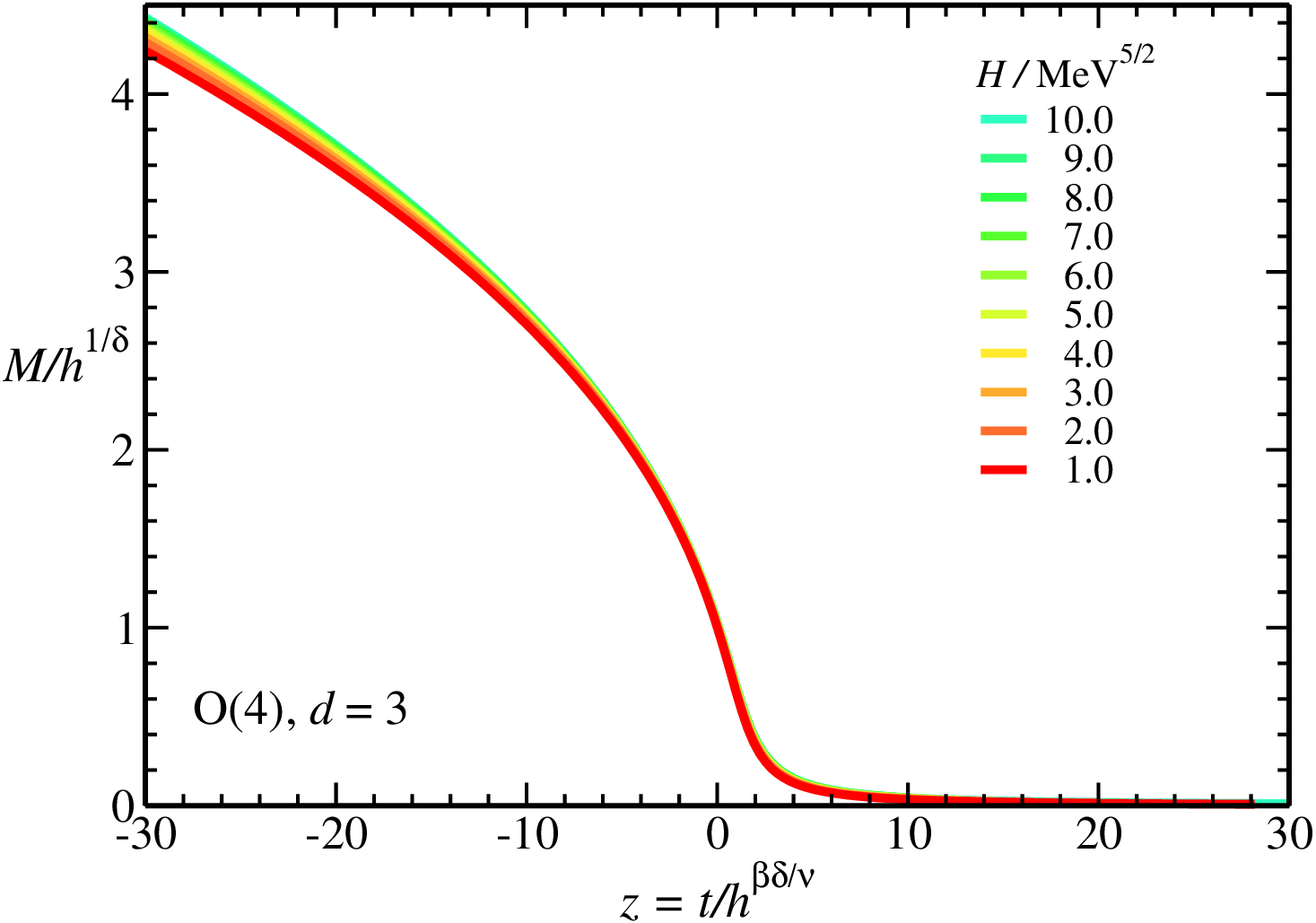}
  \caption{\label{fig:IVS}Scaling behavior of the order parameter in infinite volume. Order parameter $M$ vs. temperature $t$ for different values of the symmetry-breaking field $H$ (left panel), and scaled order parameter $M/h^{1/\delta}$ vs. the scaling variable $z = t/h^{\beta \delta/\nu}$ for the same values of the field $H$.}
\end{figure}

Close to the critical point, the order parameter $M$ satisfies the scaling relation
\begin{eqnarray}
M(t, h) = h^{1/\delta} f(z), \;\; z=t/h^{1/(\beta \delta)},
\end{eqnarray}
where $z$ is the scaling variable and $f(z)$ is a universal scaling function. The dimensionless temperature and field parameters $t=(T-T_c)/T_0$ and $h=H/H_0$ are normalized such that $M(t, h=0) = (-t)^\beta$ and $f(0)=1$.
In Fig.~\ref{fig:IVS}, results for the order parameter as a function of the temperature $t$ for different values of $H$ are plotted in the right panel. In a clear indication of scaling behavior, these curves collapse onto the universal scaling function $f(z)$ when $M/h^{1/\delta}$ is plotted as a function of $z$ (left panel). This confirms that the critical exponents are correctly determined and that the scaling behavior is captured.

\section{Finite Size-Scaling}

Because universal behavior depends on a diverging correlation length $\xi$, and because a finite volume provides a natural infrared cutoff $L$, putting a system in a box is going to influence the scaling behavior in the vicinity of a critical point. According to the finite-size scaling hypothesis \cite{Fisher:1971ks}, the scaling behavior of a thermodynamic observable depends only on the ratio between the infinite-volume correlation length $\xi$ and the size of the box $L$. 
For example, for the order parameter $M$ as a function of temperature $t$, the ratio of the order parameter $M_L(t)$ in a finite volume of size $L$ at temperature $t$ and of the order parameter $M_{\infty}(t)$ in infinite volume at the same temperature is a function of the ratio of the correlation length
 at this temperature and the volume size: 
\begin{eqnarray}
\frac{M_{L}(t)}{M_{\infty}(t)} &=& {\mathcal F}\left(\frac{\xi(t)}{L}\right).
\end{eqnarray}
Starting from this hypothesis, one can obtain universal finite-size scaling functions. 
In order to keep the ratio of the correlation length and the box size constant and thus keep the finite-size effect the same, one has to vary the temperature with the box size according to $t \sim L^{-1/\nu}$,
since the correlation length varies as $\xi \sim t^{-\nu}$. The situation is further complicated by the external symmetry breaking field $h$. In order to keep the physical behavior the same while varying the temperature, one must also keep the scaling variable $z= t/h^{1/(\beta \delta)}$ constant, and thus needs to vary $h$ according to $h \sim L^{-\beta \delta /\nu}$.
Taking into account the infinite-volume scaling behavior, e.g. for the order parameter $M(t, h) = h^{1/\delta} f(z)$, one finds that the combination
\begin{eqnarray}
L^{\beta/\nu} M &=& Q_M(z, h L^{\beta\delta/\nu}) 
\end{eqnarray}
ought to be a universal finite-size scaling function where scaling holds.
\begin{figure}
  \includegraphics[height=.22\textheight]{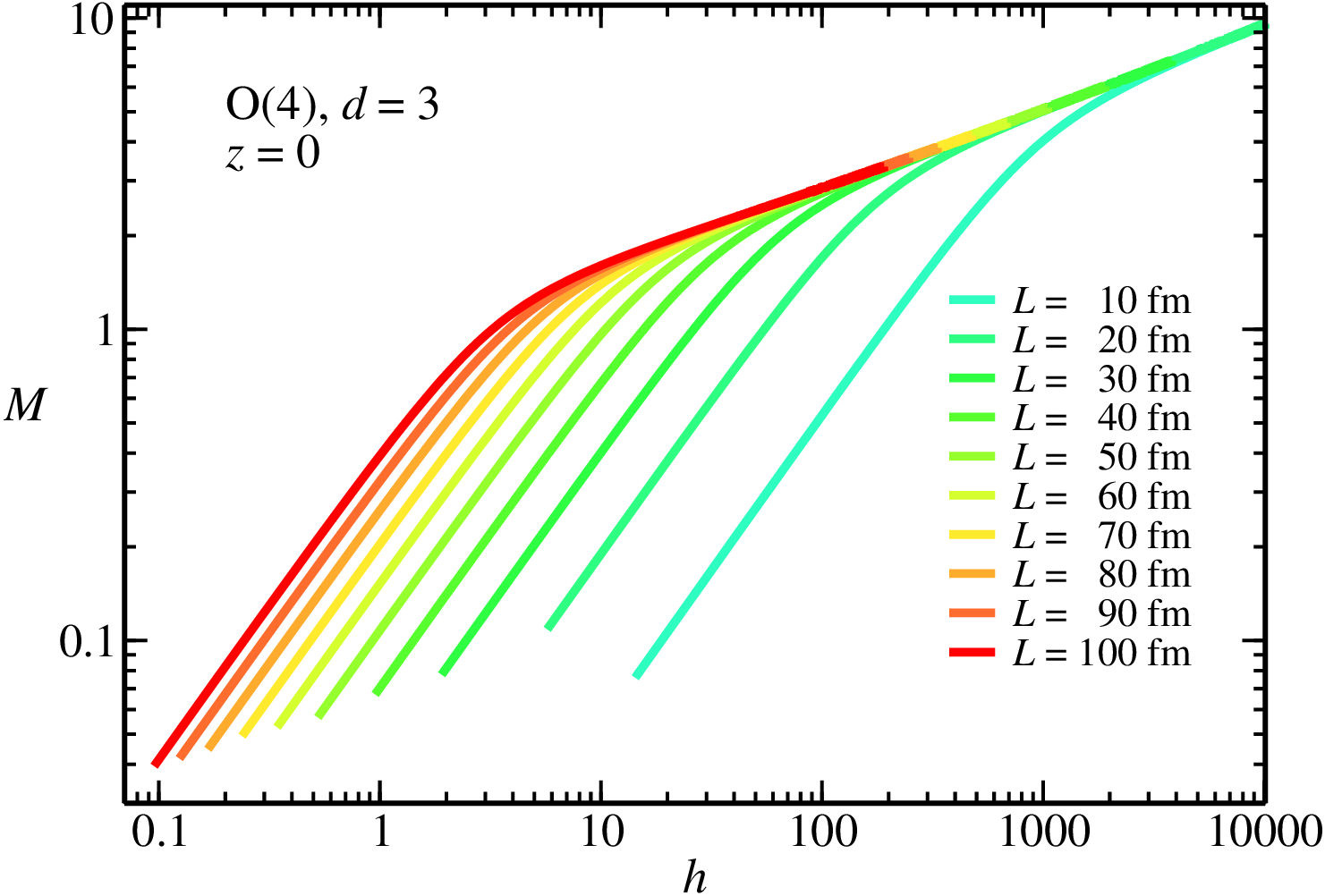}
   \includegraphics[height=.22\textheight]{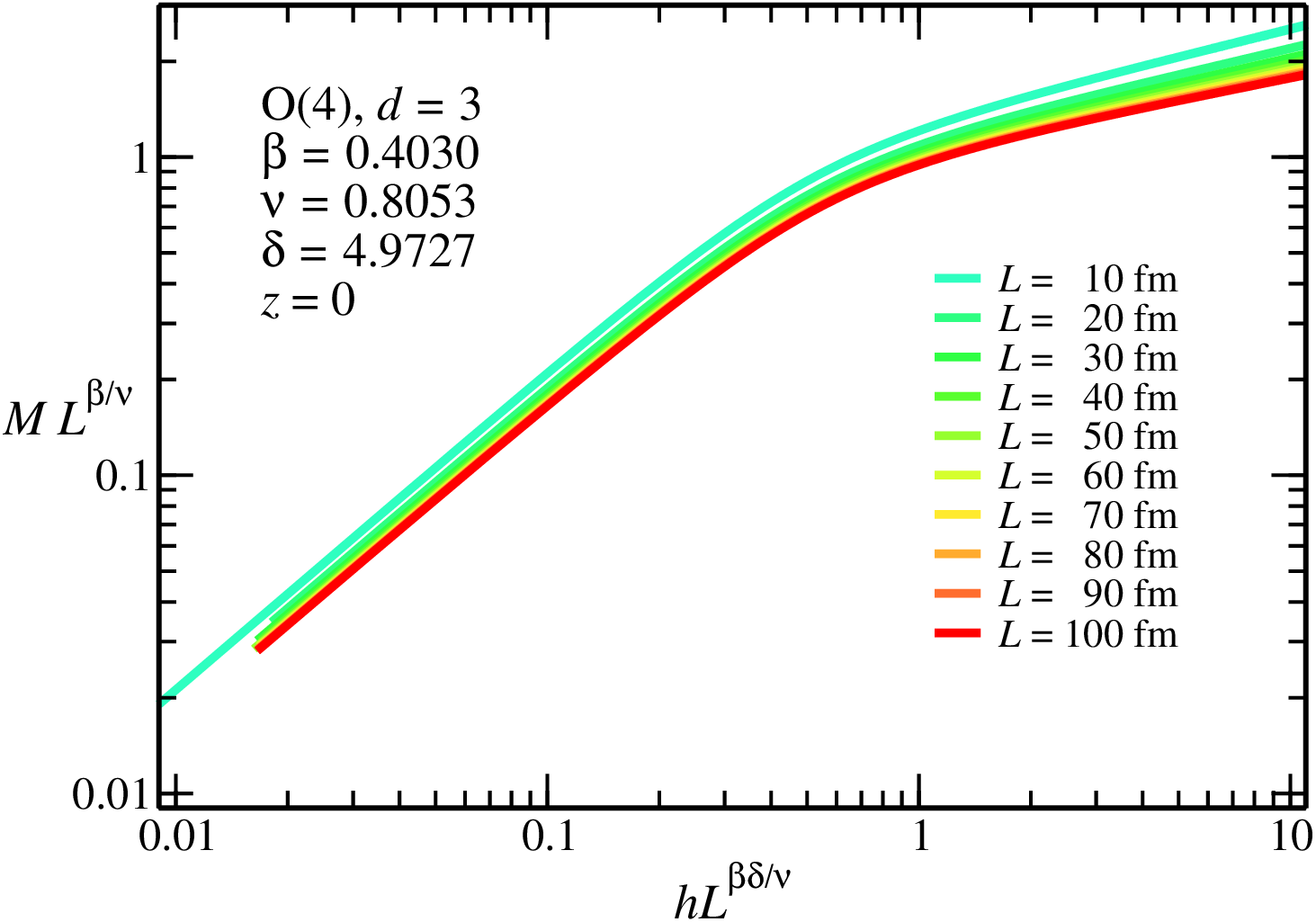}
  \caption{\label{fig:FSO}Finite-size scaling behavior of the order parameter at the critical temperature. Order parameter $M$ vs. field $h$ for different volume size $L$ (left panel), and finite-size scaled order parameter $M L^{\beta/\nu}$ vs. the finite-size scaling variable $ h L^{\beta \delta/\nu}$ for the same values of the volume (right panel).}
\end{figure}
\begin{figure}
  \includegraphics[height=.22\textheight]{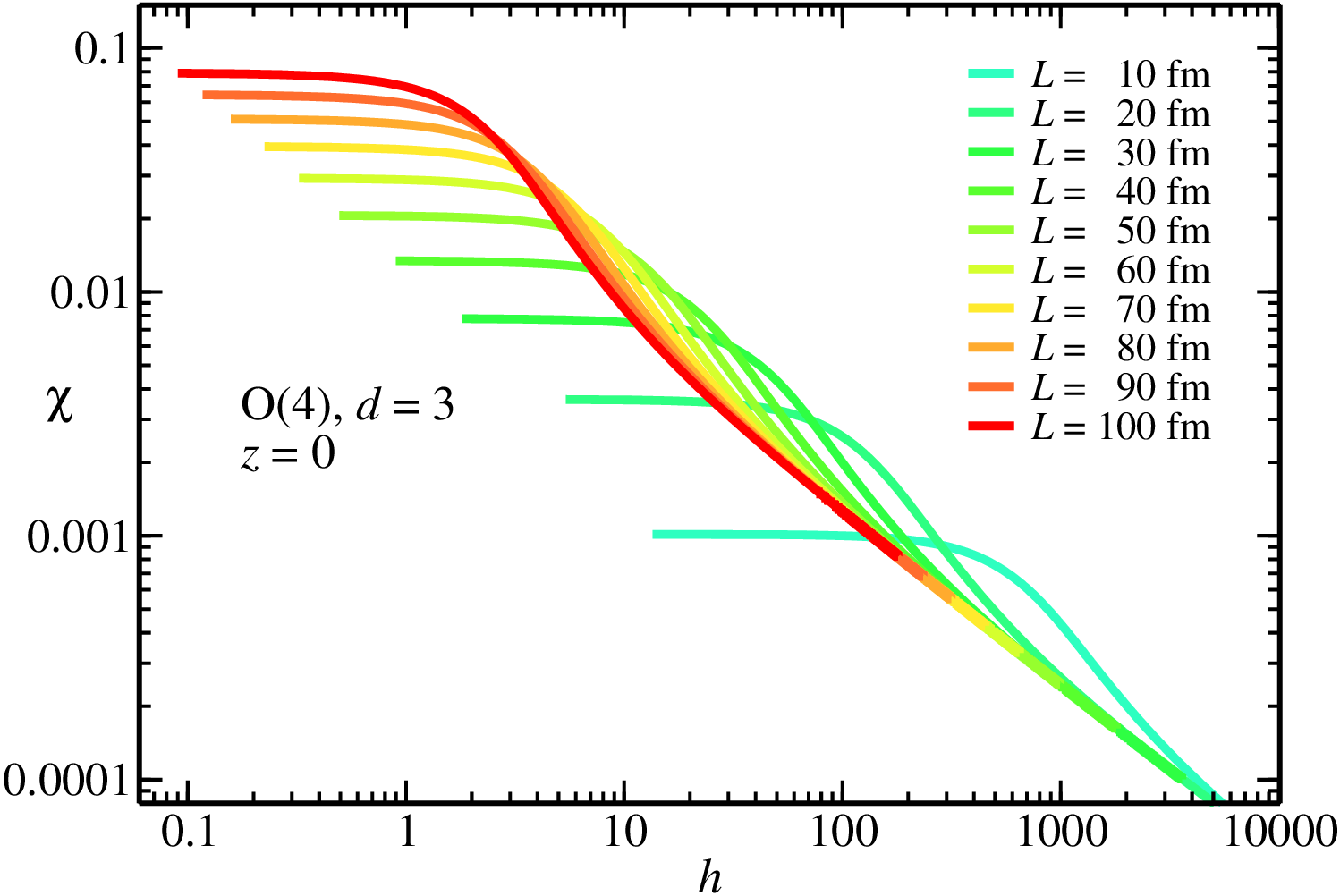}
   \includegraphics[height=.22\textheight]{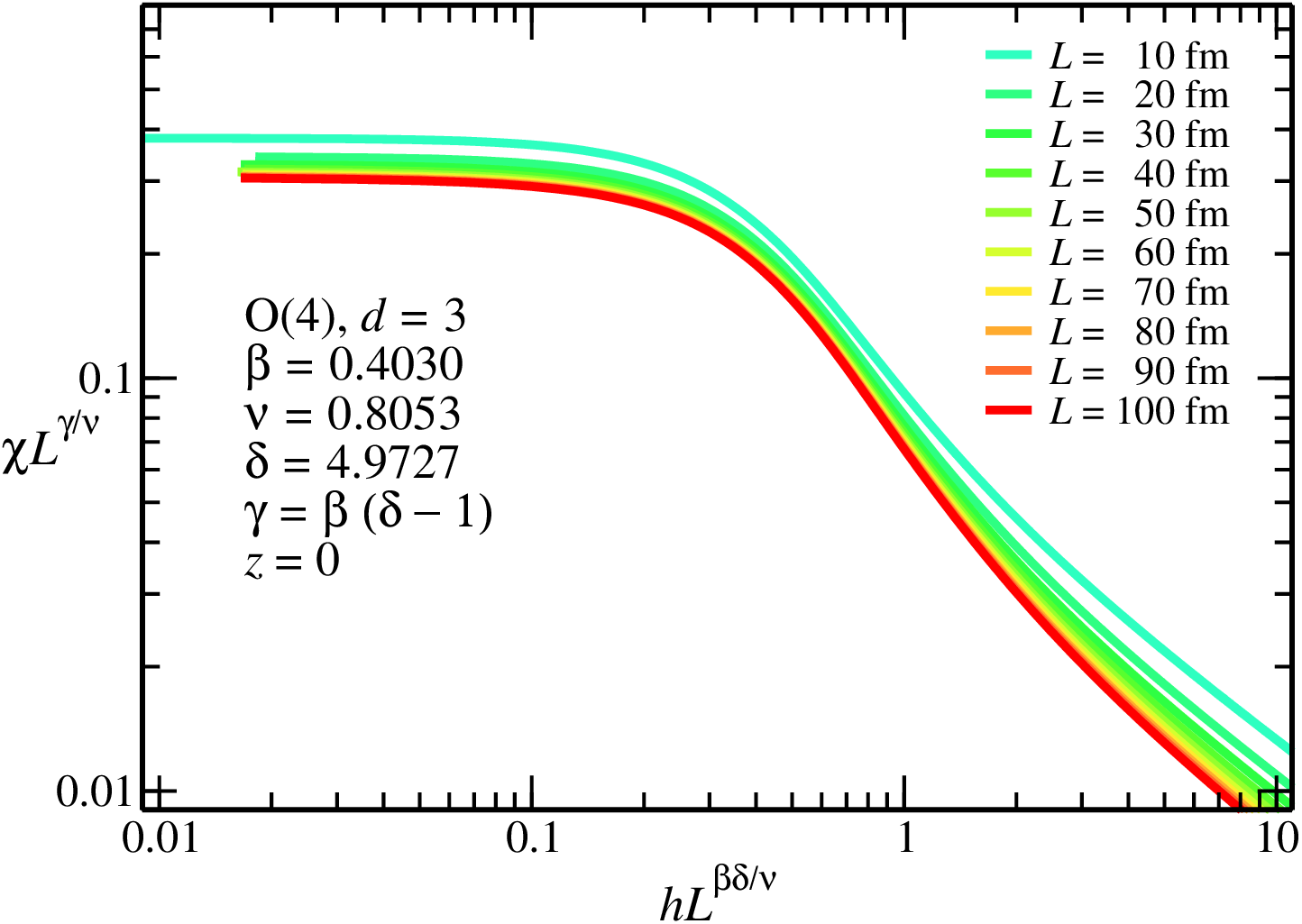}
  \caption{\label{fig:FSS}Finite-size scaling behavior of the susceptibility at the critical temperature. Susceptibility $\chi$ vs. field $h$ for different volume size $L$ (left panel), and finite-size scaled susceptibility $\chi L^{\gamma/\nu}$ vs. the finite-size scaling variable $ h L^{\beta \delta/\nu}$ for the same values of the volume (right panel).}
\end{figure}  
In Fig.~\ref{fig:FSO}, results for the order parameter $M$ as a function of the field $h$ at the critical temperature ($z=0$) are plotted for different volume sizes from $10$ to $100$ fm. In the left panel, one can clearly see the deviation from the infinite-volume scaling behavior. For large values of $h$, the mass of the fluctuations is large, the correlation length is short, and deviations from the asymptotic infinite-volume scaling behavior occur only for very small volume size. For large volume size, deviations occur for small $h$, where the correlation length is of the order of the volume size. In the right panel, the finite-volume scaled order parameter is plotted against the scaling variable, and the curves for different volume size collapse for small $h$ onto a single curve, as expected.

In Fig.~\ref{fig:FSS}, results for the susceptibility $\chi$ as a function of $h$ at the critical temperature ($z=0$) are shown. Once again, the right panel clearly shows that the finite-size scaling behavior is as expected, the curves for different volume size, which differ by two orders of magnitude, collapse onto a single curve after rescaling.
The fact that scaling works so well also validates our results for the critical exponents, and it shows that the calculation method indeed includes the long-range fluctuations responsible for scaling.

Although the results in Figs.~\ref{fig:FSO} and \ref{fig:FSS} coincide quite well, the scaling behavior is obviously not perfect. 
The deviations can be understood by analyzing the scaling corrections to the leading behavior.
From an RG analysis, one can show that the corrections, which depend explicitly on the volume, are of the form
\begin{eqnarray}
L^{\beta/\nu} M &=&  Q_{M}^{(0)}(z, h L^{\beta\delta/\nu}) + \frac{1}{L^\omega}  Q_{M}^{(1)}(z, h L^{\beta\delta/\nu}) + \ldots 
\label{eq:corr}
\end{eqnarray}
The exponent $\omega$ is associated with the first irrelevant RG operator at the critical fixed point. The coefficient functions depend only on the scaling variables $z$ and $h L^{\beta \delta /\nu}$.
From the deviations, we determine $\omega = 0.74(4)$.  This in good agreement with the result $\omega = 0.7338$ of the fixed point analysis in the same RG scheme in \cite{Litim:2001hk}.

The main conclusion from the scaling corrections is that it might not be enough to analyze the leading-order scaling behavior, in particular for very  small volumes. In the present case, where the scale 
is set by a UV cutoff of the order of  a typical thermodynamic lattice cutoff, deviations are already sizable at $L=10$ fm. For the O(2) class, where the scaling region is narrow, a careful analysis is warranted.

\section{Conclusions}

In this contribution, we address the question of finite-size scaling in the O(N) universality class with non-perturbative Renormalization Group methods. Critical exponents and the universal scaling functions for the O(4) universality class in $d=3$ dimensions in infinite volume have been obtained. The scaling behavior validates our results for the critical exponents and shows that our RG scheme accounts for long-range fluctuations.

We have further demonstrated finite-size scaling behavior by implementing the RG scheme in a finite volume. We have calculated the universal finite-size scaling functions for the order parameter $M$ and the susceptibility $\chi$ for a wide range of values for the scaling variable $z$. Deviations from the leading-order scaling behavior are found to be consistent with the expected corrections from irrelevant operators. Due to these scaling corrections, an analysis taking into account only the leading scaling behavior might not be sufficient for small volumes used in current lattice simulations.

Our results are suitable for a direct comparison to results from lattice QCD to check compatibility of the scaling behavior with the O(4) universality class. We plan to carry out such a comparison, and will also extend our results to the O(2) class. 


\begin{theacknowledgments}
BK would like to thank the organizers for a very enjoyable workshop. Useful discussions with A. Di Giacomo, M. D'Elia and G. Pica,  and with T. Mendes are gratefully 
acknowledged. This work was supported in part by the Excellence cluster "Origins and structure of the Universe", and by the Natural Sciences and Engineering Research Council of Canada (NSERC).  TRIUMF receives federal funding via a contribution agreement through the National Research Council of Canada.
\end{theacknowledgments}



\bibliographystyle{aipproc}   

\bibliography{klein_bertram_fsscaling}

\IfFileExists{\jobname.bbl}{}
 {\typeout{}
  \typeout{******************************************}
  \typeout{** Please run "bibtex \jobname" to optain}
  \typeout{** the bibliography and then re-run LaTeX}
  \typeout{** twice to fix the references!}
  \typeout{******************************************}
  \typeout{}
 }

\end{document}